\documentclass[journal=nalefd,manuscript=article]{achemso}

\usepackage[version=3]{mhchem} 
\usepackage{xcolor}

\usepackage{graphicx}
\usepackage{color}
\usepackage{dcolumn}
\usepackage{amsmath}
\usepackage{amsfonts}
\usepackage{bm}
\usepackage{epstopdf}
\usepackage{picture}
\usepackage{enumitem}
\usepackage{afterpage}
\usepackage{placeins}
\usepackage{float}
\usepackage{ulem}
\usepackage{xr}

\author{Paolo Polimeno}
\affiliation[CNR-IPCF]{CNR-IPCF, Istituto per i Processi Chimico-Fisici, I-98158, Messina, Italy}
\alsoaffiliation[Universit\`{a} di Messina] {Dipartimento di Scienze Matematiche e Informatiche,
Scienze Fisiche e Scienze della Terra, I-98166, Universit\`{a} degli Studi di Messina, Italy}
\author{Francesco Patti}
\affiliation[CNR-IPCF]{CNR-IPCF, Istituto per i Processi Chimico-Fisici, I-98158, Messina, Italy}
\alsoaffiliation[Universit\`{a} di Messina] {Dipartimento di Scienze Matematiche e Informatiche,
Scienze Fisiche e Scienze della Terra, I-98166, Universit\`{a} degli Studi di Messina, Italy}
\author{Melissa Infusino}
\affiliation[USFQ]{Colegio de Ciencias e Ingenieria, Universidad San Francisco de Quito, Ecuador}
\author{Jonathan S\'{a}nchez}
\affiliation[USFQ]{Colegio de Ciencias e Ingenieria, Universidad San Francisco de Quito, Ecuador}
\author{Maria A. Iat\`{\i}}
\affiliation[CNR-IPCF]{CNR-IPCF, Istituto per i Processi Chimico-Fisici, I-98158, Messina, Italy}
\author{Rosalba Saija}
\affiliation[Universit\`{a} di Messina] {Dipartimento di Scienze Matematiche e Informatiche,
Scienze Fisiche e Scienze della Terra, I-98166, Universit\`{a} degli Studi di Messina,
Italy}\alsoaffiliation[CNR-IPCF]{CNR-IPCF, Istituto per i Processi Chimico-Fisici, I-98158,
Messina, Italy}
\author{Giovanni Volpe}
\affiliation[UniGoteborg]{Institutionen f{\"o}r Fysik, G{\"o}teborgs Universitet, I-41296,
G{\"o}teborg, Sweden}
\author{Onofrio M. Marag\`{o}}
\email{onofrio.marago@cnr.it}\affiliation[CNR-IPCF]{CNR-IPCF, Istituto per i Processi
Chimico-Fisici, I-98158, Messina, Italy}
\author{Alessandro Veltri}
\email{aveltri@usfq.edu.ec}\affiliation[USFQ]{Colegio de Ciencias e Ingenieria, Universidad San
Francisco de Quito, Ecuador}

\title[Gain-Assisted Optomechanical Locking of Metal/Dielectric Nanoshells in Optical Potentials]{Gain-Assisted Optomechanical Position Locking of Metal/Dielectric Nanoshells in Optical Potentials}


\begin{document}

\begin{abstract}
We investigate gain-assisted optical forces on dye-enriched silver nanoshell in the quasi-static
limit by means of a theoretical/numerical approach. We demonstrate the onset of nonlinear optical
trapping of these resonant nanostructures in a counter-propagating Gaussian beam configuration. We
study the optical forces and trapping behaviour as a function of wavelength, particle gain level,
and laser power. We support the theoretical analysis with Brownian dynamics simulations that show
how particle position locking is achieved at high gains in extended optical trapping potentials.
Finally, for wavelengths blue-detuned with respect to the plasmon-enhanced resonance, we observe
particle channeling by the standing wave antinodes due to gradient force reversal. This work opens
perspectives for gain-assisted optomechanics where nonlinear optical forces are finely tuned to
efficiently trap, manipulate, channel, and deliver externally controlled nanophotonic system.
\end{abstract}


Optical tweezers\cite{Ashkin1986,jones2015optical} (OT) are crucial tools for the manipulation and
study of micro- and nanoscopic particles of different nature without mechanical
contact\cite{JonasEPH08,polimeno2018optical}. In recent years, a tremendous effort has been devoted
to the optical trapping and optical manipulation of nanoparticles in liquid, air or
vacuum\cite{Marago2013,Spesyvtseva2016}. The difficulties in optical trapping nano-sized matter are
mainly related to the fact that optical forces decrease with the particle volume for small
particles\cite{polimeno2018optical}, which yields trapping potentials lower than the energy of
thermal fluctuations for reasonable incident laser powers\cite{Marago2013}. Standard OT,
\textit{i.e.}, single-beam optical traps, are also affected by the unavoidable light scattering
forces which tend to push the particle along the light propagation direction, and might have a
particularly destabilizing effect for highly absorbing, resonant or plasmonic
nanoparticles\cite{Marago2013,Spesyvtseva2016}. Morphology\cite{YanACSNANO13b,Irrera2016}, material
composition\cite{VanderHorst2008,Marago2008a,Donato2018}, material hybridization\cite{Spadaro2015,
Spadaro2016}, and resonant opto-plasmonic response\cite{Lehmuskero2015,BrzobohatySR15,Messina2015}
can increase optical trapping at the nanoscale. Scattering forces can be balanced at equilibrium in
dual-beam optical traps based on the use of low numerical aperture lenses in a counterpropagating
beam geometry\cite{Ashkin1970}. For laser beams  with the same polarization, a standing wave is
formed with an intensity modulation along the beam axis that generate an optical potential with
many equilibrium
position\cite{ZemanekJOSAA02,CizmarAPB06,GherardiAPL08,Singer2003,donato2019optical}.

Among the various applications of optical forces at the nanoscale, the study of optical forces in
optically trapped gain-enriched plasmonic nanostructures appears to be of particular interest. In
fact, a plethora of remarkable phenomena occurs in these
systems\cite{veltri2012optical,infusino2014loss} due to the resonant interplay between plasmonic
structures and gain media (\textit{e.g.} dye molecules or quantum dots). In particular, the
coupling with a gain medium located in the core of a metallic nanoshell, when excited by means of
an external pump, produces intense changes of the electromagnetic fields around the structure, thus
producing novel features which can be useful for a variety of applications, such as photothermal
therapy, enhanced spectroscopy, and
spasing\cite{caligiuri2016resonant,veltri2016multipolar,pezzi2019resonant}.

Here, we present a study of the optical forces acting on a gain-enriched silver nanoshell in the
quasi-static limit. Specifically, we analyze the optomechanical response of this nanostructure in a
counterpropagating Gaussian beam optical trap, where a systematic analysis can be performed without
any detrimental effect of the scattering force component. In particular, we study the behaviour of
the optical force constants as a function of wavelength and for different gain levels, which can be
achieved by fixing the molecular density of the gain medium and varying the power of the external
pump. We show that optical trapping strongly depends both on the wavelength and on the gain level.
Moreover, we investigate the stable configurations and particle dynamics in the trap by means of
Brownian dynamics simulation. Interesting localization effects appear for wavelengths red-detuned
with respect to the gain-enhanced resonance, while for blue-detuned wavelengths, we observe
particle channeling by the standing wave antinodes due to the reversal of the gradient force.

\begin{figure}
\includegraphics[width=\textwidth]{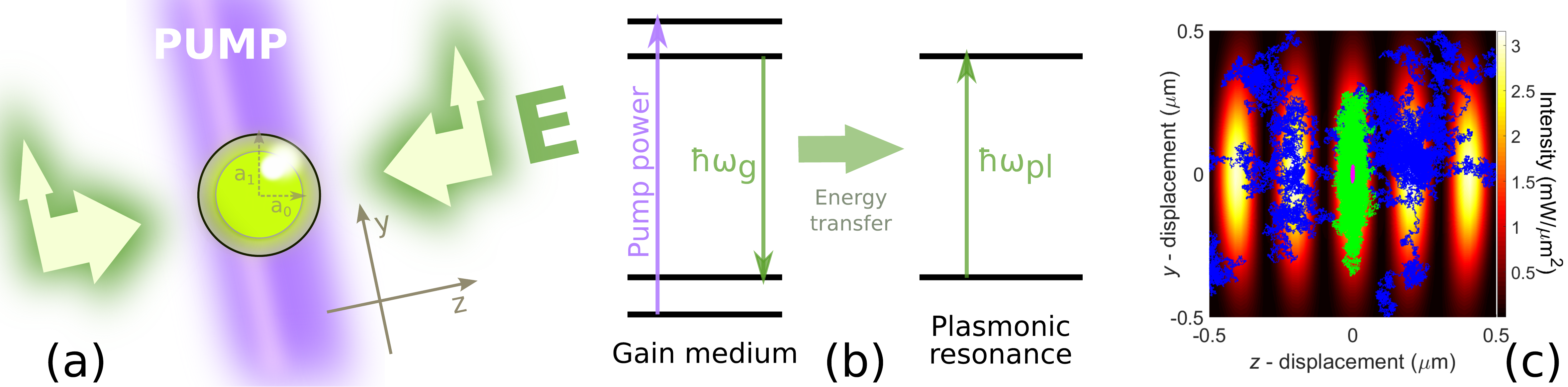}
    \caption{In (a), a schematic of the system under study is represented. A core-shell particle is composed by an external silver shell and a silica core,
    doped with rhodamine dye molecules which act like gain material under the action of a pumped field (purple) considered in a counterpropagating configuration to balance radiation pressure. This system is placed in a double-beam optical tweezers configured
    with co-linearly polarized counterpropagating Gaussian beams (green). The double beam configuration ensures the balance of scattering forces in the trap region. In (b), the scheme of the energy levels is presented. System pumping inverts the population
    $\tilde{N}$ of the dye using a pump frequency higher than the frequency of the nanoshell plasmonic resonance. In this way, an emission of
    the dye at the wavelength of the silver plasmon resonance is induced. In (c), counter-propagating laser beam intensity is represented and
    projected on the $yz$ plane. The probe wavelength is $\lambda=531.9$ nm, its single-beam
    power is 50 mW, and NA$= 0.6$. Nanoparticle trajectories is plotted for
    $G = 0$ (blue), $G = - 0.132$ (green), and $G = - 0.22$ (magenta). Note how the highest gain (magenta) shows the tightest confinement. Brownian dynamics simulations are performed for all cases with a time step of $\Delta t=10^{-7}$ s and a sampling time of $t_\mathrm{samp} =0.1$ s.}
  \label{figure1}
\end{figure}

\section{Theoretical model}
\paragraph{Metal/dielectric nanoshells.}
The system under study (Fig.~\ref{figure1}) is a metal nanoshell embedding a gain enriched
dielectric core, studied in the quasi-static limit and below the emission
threshold\cite{veltri2016multipolar} (\textit{i. e.}, when the gain is not enough to completely
overcome the metal losses). In this regime, the nanoshell geometry  ensures that the plasmonic
field can be described as dipolar without any approximation.
 Thus, we can
use a steady state model for the particle polarizability $\alpha$:
\begin{equation}\label{eq:polarizabily}
\alpha=\frac{(\varepsilon_\mathrm{sh}-\varepsilon_\mathrm{m})
(\varepsilon_\mathrm{h}+2\varepsilon_\mathrm{sh}) +
f^3(\varepsilon_\mathrm{h}-\varepsilon_\mathrm{sh})
(\varepsilon_\mathrm{m}+2\varepsilon_\mathrm{sh})}
{(\varepsilon_\mathrm{h}+2\varepsilon_\mathrm{sh})
(\varepsilon_\mathrm{sh}+2\varepsilon_\mathrm{m}) +
2f^3(\varepsilon_\mathrm{h}-\varepsilon_\mathrm{sh})
(\varepsilon_\mathrm{sh}-\varepsilon_\mathrm{m})} ,
\end{equation}
where $\varepsilon_\mathrm{sh}(\omega)$ is the complex steady state metal shell permittivity,
$\varepsilon_\mathrm{h}(\omega)$ is the complex steady state permittivity of the gain enriched
core, $\varepsilon_\mathrm{m}$ is a real number representing the permittivity of the  external
medium (solvent) hosting the nanoparticle,
 and $f=a_0/a_1$ is the ratio between the internal and the external radius of the nanoshell (Fig.~\ref{figure1}a).
In all of the calculations and simulations presented in this work the nanoparticle has an external
radius of $a_1=20$~nm and a ratio between the internal and the external radius of $f=0.8$ (meaning
that the internal radius is $a_0=16$~nm).

For the metal permittivity we used a Drude-Lorentz model in which the losses due to the interband
transitions $\delta\varepsilon_\mathrm{sh}(\omega)$ have been added heuristically by interpolating
them from the Johnson and Christie data set\cite{johnson1972optical}{}:
\begin{equation}
 \varepsilon_\mathrm{sh}(\omega)=\varepsilon_\infty+\delta\varepsilon_\mathrm{sh}(\omega)-\frac{\omega_{\mathrm{pl}}^2}{\omega(\omega+2i\Gamma)}
 ,
\end{equation}
where $\varepsilon_\infty$ is a constant offset for the real part of the dielectric function,
accounting for the high frequency contributions and the vacuum contribution, $\omega_\mathrm{pl}$
is the plasma frequency and $\Gamma$ is the ionic collisions friction coefficient. In all of our
calculations we will use $\varepsilon_\infty=5.3$, $\hbar\omega_\mathrm{pl}=9.6$~eV and
$\hbar\Gamma=0.0228$~eV which are compatible with the complex dielectric function of silver. As one
can see in Fig.~\ref{figure1}b, the gain elements (\textit{e.g.}, molecule/quantum dot) have been
chosen in order to resonate with the plasmon. Being below the emission threshold we can model that
using a Lorentzian shape:
\begin{equation}
    \varepsilon_\mathrm{h}(\omega)= \varepsilon_\mathrm{b}-\frac{G\Delta}{2(\omega-\omega_\mathrm{g})+i\Delta}
    ,
\end{equation}
where $\varepsilon_\mathrm{b}$ is the permittivity of the dielectric core in which the gain
elements are embedded, $\omega_\mathrm{g}$ is the emission centerline of the gain elements,
$\Delta=2/\tau$ is the width of the Lorentzian shape where $\tau$ is the time constants associated
with energy (spontaneous emission) relaxation processes of the gain element, and $G$ is a
dimensionless parameter equal to the maximum in absolute value of the imaginary part of
$\varepsilon_\mathrm{h}(\omega)$ and measuring the amount of gain present in the system; it can be
shown to be equal to\cite{veltri2016multipolar}:
\begin{equation}\label{epsm}
 G = \Im[\varepsilon_\mathrm{h}(\omega_g)] =-\frac{n\mu^2\tau}{3\hbar\varepsilon_0}\tilde{N},
\end{equation}
where $n$ is the element density of the gain medium, $\mu$ is the amplitude of the transition
dipole moment of the gain element and $\tilde{N}$ represents the population inversion produced
through the action of the external pump (\textit{e.g.}, when $\tilde{N}=0$ all of the gain
molecules are in the ground state, when $\tilde{N}=1$ they are in the excited state), $\hbar$ is
the reduced Planck constant and $\varepsilon_0$ is the vacuum permittivity. For a given gain
molecule (where $\mu$, $\omega_g$ and $\tau$ are set), the molecule density $n$ describes the
maximum possible value for $G$ although, even once the molecule density $n$ of the nanoparticle's
core is fixed, $G$ can be experimentally modulated by means of the external pump power, from zero
(corresponding to $\tilde{N}=0$ or pump off) to its maximum value (corresponding to $\tilde{N}=1$
which is when the system is completely pumped). In all of our simulations, $G$ ranged between zero
and $G=-0.22$; this last minimum value, when using conservative estimates for the physical
quantities in Eq.~\ref{epsm} (such as a transition dipole moment of $\mu=10$~D and a relaxation
time $\tau=10^{-14}$~s) corresponds to a element density of $n=0.35$~nm$^{-3}$. It is worth noting
that this estimation has been done for the {\it minimum value} we used for $G$ (corresponding to
the the highest gain), while the effect on the optical forces begin to appear for much lower gain
levels. This means that, even if we are neglecting  efficiency-reducing quenching effects due to
the proximity of the gain element to the metal in the nanoparticle, we are working in a realistic
range of gain elements density.

\paragraph{Gain-assisted nonlinear optical trapping.}
We consider a silver nanoshell with an external radius of $20$ nm and trapping light with a
wavelength in the visible range. Therefore, the particle size is small enough that optical forces
can be calculated within a dipole
approximation\cite{chaumet2000time,arias2003optical,albaladejo2009scattering} and expressed in
terms of the linear effective complex polarizability, $\alpha$, that in our model is given by
Eq.~\ref{eq:polarizabily}. It is worth noting that the trapping light provides also the probe field
exciting the plasmonic resonance of the nanoshell, meaning that the dipole moment $\bf p$ of the
nanoparticle can be written as:
\begin{equation}
{\bf p}=\alpha{\bf E},
\end{equation}
where $\bf E$ is the electric field associated with the trapping light. Thus, the time-averaged
optical force experienced by the nanoshell when illuminated by the incident light
is\cite{gao2017optical,polimeno2018optical}:
\begin{equation}
\mathbf{F}_{\mathrm{DA}}=\frac{1}{2} \Re \left\{\sum_i \alpha E_i \nabla E_i^{\ast} \right\},
\end{equation}
\noindent where $E_i$ are the electric field components. Starting from this expression, one can
explicitly write the optical force in terms of extinction cross-section $\sigma_{\mathrm{ext}}$ and
particle's polarizability\cite{jones2015optical}:

\begin{equation}\label{eq:forcesDA}
\mathbf{F}_{\mathrm{DA}}(\textbf{r}) = \frac{n_\mathrm{m} \Re \left\{ \alpha \right\}}{2 c
\varepsilon_0 \varepsilon_{\mathrm{m}}} \nabla I(\textbf{r}) + \frac{n_\mathrm{m}
\sigma_\mathrm{ext}}{c} \mathbf{S},
\end{equation}

\noindent where $n_\mathrm{m}$ is the medium refractive index (water in our calculation,
$n_\mathrm{m}=1.33$), $c$ is the light velocity, $I(\textbf{r}) = n_\mathrm{m} c \vert
\mathbf{E}(\textbf{r}) \vert^{2}/2$ is the intensity of the electric field, and $\mathbf{S} = \Re
\{ \mathbf{E} \times \mathbf{H}^{*} \}/2$ is the time-averaged Poynting vector of the incoming
wave\cite{gao2017optical,albaladejo2009scattering} that is related to the incident light intensity,
$|\mathbf{S}|=I(\textbf{r})$. The first term in Eq.~\ref{eq:forcesDA} represents the gradient force
and it is responsible for particle confinement in optical tweezers. Since it arises from the
potential energy of the induced dipole immersed in the electric field, it is conservative.
Particles with a positive $\Re\left\{\alpha\right\}$ are attracted towards the high intensity
region of the optical field. Conversely, when the real part of the polarizability is negative the
particles are repelled by the high intensity region. The second term in Eq.~\ref{eq:forcesDA} is
the scattering force. It is responsible for the radiation pressure, it is not conservative and it
is directed along the direction of propagation of the laser beam \cite{Ashkin1970}.

We now focus on the modeling of the double-beam OT in their standing wave
configuration\cite{zemanek1998optical} (see the Supplementary Information for a detailed analysis
for single-beam OT). This is realized with counterpropagating Gaussian beams and a low numerical
aperture objective that we fix at a typical value\cite{BrzobohatySR15,donato2019optical} of
NA$=0.6$. The two counterpropagating Gaussian beams are considered in a paraxial
approximation\cite{jones2015optical}, they propagate along the $z$ axis and their waists coincide
with the origin of the laboratory reference system (see Fig.~\ref{figure1}a). Moreover, the
polarization directions of the two beams are co-linear and lie on the $xy$ plane. In this manner,
the total light intensity, that is a function of the radial, $\rho$, and axial, $z$, directions,
assumes a standing wave profile\cite{zemanek1998optical,bernatova2019wavelength}:
\begin{equation}\label{eq:intensityDA}
  I(\rho,z) = \frac{4 I_0 w_0^2}{w^2(z)} e^{\frac{-2 \rho^2}{w^2(z)}} \cos^2 \Phi(z),
\end{equation}
where $I_0 = 2 P/\pi {w_0}^2$ is the maximum intensity, $P$ is the single Gaussian laser beam
power, $w_0 = 0.5 \lambda/\mathrm{NA}$ is the beam waist which is evaluated with the Abbe
criterion, $\lambda$ is the wavelength in vacuum, $w(z)$ is the beam width such that $w(z) = w_0
\sqrt{1+z^2/z_0^2}$, $\Phi(z) = k_\mathrm{m}z - \zeta(z) + k_{m}\rho^2 / 2 R(z)$, $R(z)$ is the
wavefront radius $R(z) = z ( 1 + z^2/z_0^2 )$, $\zeta(z)$ is phase correction $\zeta(z) =
\mathrm{atan}( z/z_0 )$ and $z_0$ is the Rayleigh range which denotes the distance from the beam
waist at $z=0$ to where the beam width has increased by a factor $\sqrt{2}$, $z_0 = k_\mathrm{m}
w_0^2/2$. In Eq.~\ref{eq:intensityDA}, the interference between the two Gaussian beams generates a
standing wave with a modulation of intensity along the $z$-axis that results in a strong
wavelength-dependent modulation of the axial optical force. In Fig.~\ref{figure1}c, a quantitative
representation of the standing wave intensity, as projected in $yz$ plane, is provided together
with the nanoshell tracking obtained by Brownian dynamics simulation (see below). Thus, using
Eq.~\ref{eq:forcesDA} and Eq.~\ref{eq:intensityDA}, we get the expression of the gradient force
components\cite{zemanek1998optical}:

\begin{equation}\label{eq:forceDA_CP}
    \begin{split}
    \langle F \rangle&_{\mathrm{DA},\rho}(\rho,z) = - \frac{4 \Re\left \{\alpha\right\} I_0 w_0^2  \rho e^{\frac{-2 \rho^2}{w^2(z)}}}{c \varepsilon_0 n_\mathrm{m} w^4(z)} \left( \cos^2 \Phi(z) + \frac{z \sin 2 \Phi(z)}{k_\mathrm{m}w^2(z)} \right) \\
    \langle F \rangle&_{\mathrm{DA},z}(\rho,z) = - \frac{4 \Re\left \{\alpha\right\} I_0 w_0^2 e^{\frac{-2 \rho^2}{w^2(z)}}}{c \varepsilon_0 n_\mathrm{m} k_\mathrm{m}^2 w^2(z)} \\
    &\cdot \Bigg[\left( 1 - \frac{2 \rho^2}{w^2(z)} \right) \frac{2 z \cos^2 \Phi(z)}{w_0^2 w^2(z)} + \left( \frac{k_\mathrm{m}^2}{4} - \frac{1}{2 w^2(z)} - \rho^2 \frac{w^2(z) - 2 w_0^2}{w_0^2 w^4(z)} \right) k_\mathrm{m}  \sin 2 \Phi(z)
    \Bigg].
    \end{split}
\end{equation}

\begin{figure}
\centering
\includegraphics[width=\textwidth]{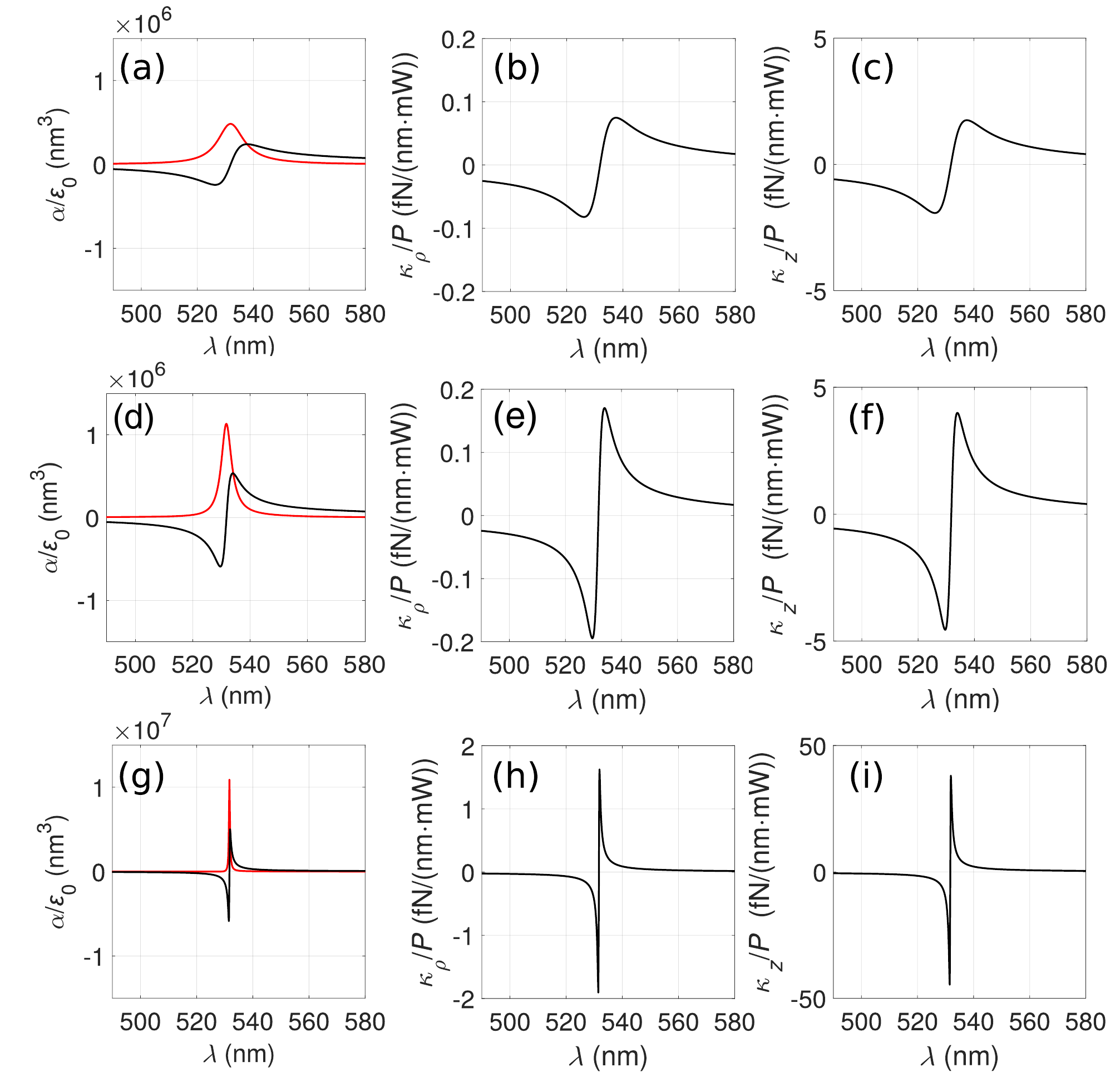}
\caption{Trap stiffnesses normalized to the beam power in counterpropagating configuration are
shown for three different gains.  (a-c) G = 0, (d-e) G = -0.132, and (g-i) G = -0.22. The first
column represent the real (black line) and imaginary (red line) parts of the polarizability
normalized by $\varepsilon_0$. The second and third column represent the radial and axial trap
stiffnesses, $\kappa_\rho$ and $\kappa_z$, normalized by the optical power, respectively.}
\label{figure2}
\end{figure}

\section{Results}
We start our analysis of the nanoshell trapping behaviour by studying the spring constants
associated to small displacements from the equilibrium trapping point, where the gradient force can
be approximated by a harmonic force. Therefore, in the $xy$ transverse plane ($z=0$), the radiation
intensity (Eq.~\ref{eq:intensityDA}) can be approximated as $I(\rho) \approx 4 I_0 \left( 1 -
2\rho^2/w_0^2 \right)$ so that the radial component of the gradient force in the
Eqs.~\ref{eq:forceDA_CP} can be rewritten as $\langle F \rangle_{\mathrm{DA},\rho}(\rho) =
-\kappa_{\rho}\rho$, where the trap stiffness is:
\begin{equation}\label{eq:DA_trap_stiffness_rho}
  \kappa_\rho = \frac{8 \Re\left \{\alpha\right\} I_0}{c \varepsilon_0 n_\mathrm{m} w_0^2} .
\end{equation}

Similarly, along $z$ axis ($\rho=0$), $I(z) \approx 4 I_0 \left[1 - \left( 2 z^2/z_0^2 \right)
\left( 1 - k_\mathrm{m}z_0 + k_\mathrm{m}^2 z_0^2/2 \right) \right]$, so that the axial component
of the gradient force in Eqs.~\ref{eq:forceDA_CP} is $\langle F \rangle_{\mathrm{DA},z}(z) =
-\kappa_{z}z$ with an associated spring constant:
\begin{equation}\label{eq:DA_trap_stiffness_z}
\kappa_z = \frac{8 \Re \left\{\alpha\right\} I_0}{c \varepsilon_0 n_\mathrm{m} z_0^2} \left( 1 -
k_\mathrm{m}z_0 + \frac{k_\mathrm{m}^2z_0^2}{2} \right) .
\end{equation}
In Fig.~\ref{figure2} we compare the particle polarizability for three different gain values
($G=[0;- 0.132;- 0.22]$) with the wavelength dependence of the trap constants normalized to the
power, $\kappa_\rho/P$ and $\kappa_z/P$. We can note that $\Im \left\{\alpha\right\}$ (red lines in
Fig.~\ref{figure2}a,d,g) increases for higher gain around the plasmon resonance, and its profile is
mostly peaked and narrow. As the gain increases the enhanced emission of the pumped dye, which is
suitably tuned to the frequency $\omega_\mathrm{pl}$, dominates the optical response in intensity
with respect to that of the plasmon mode. Here we have that $ \Im
\left\{\alpha\right\}(G=-0.22)/\Im \left\{\alpha\right\}(G=0) \simeq 50$ and the absorption appears
increasingly spectrally confined around the resonance. On the other hand, $\Re
\left\{\alpha\right\}$ changes sign in correspondence to the plasmonic resonance and for increasing
gain shows a sharpening of the dispersive curve. Thus, from Eqs.~\ref{eq:DA_trap_stiffness_rho},
\ref{eq:DA_trap_stiffness_z}, trap stiffnesses have a similar trend as shown in
Figs.~\ref{figure2}b, \ref{figure2}c, \ref{figure2}e, \ref{figure2}f, \ref{figure2}h,
\ref{figure2}i. In all cases we have that: i) for wavelengths lower than the resonance, the force
constants are negative, optical forces are repulsive, and the nanoshell is pushed away from the
high intensity region; ii) for wavelengths higher than the resonance, the force constants are
positive, and the nanoshell is attracted to the high intensity region of the standing wave. As
expected for a standing wave configuration, for a fixed gain, the axial force constants,
$\kappa_z$, are greater, in modulo, than the transverse ones, $\kappa_\rho$.
\begin{figure}
\centering
\includegraphics[width=\textwidth]{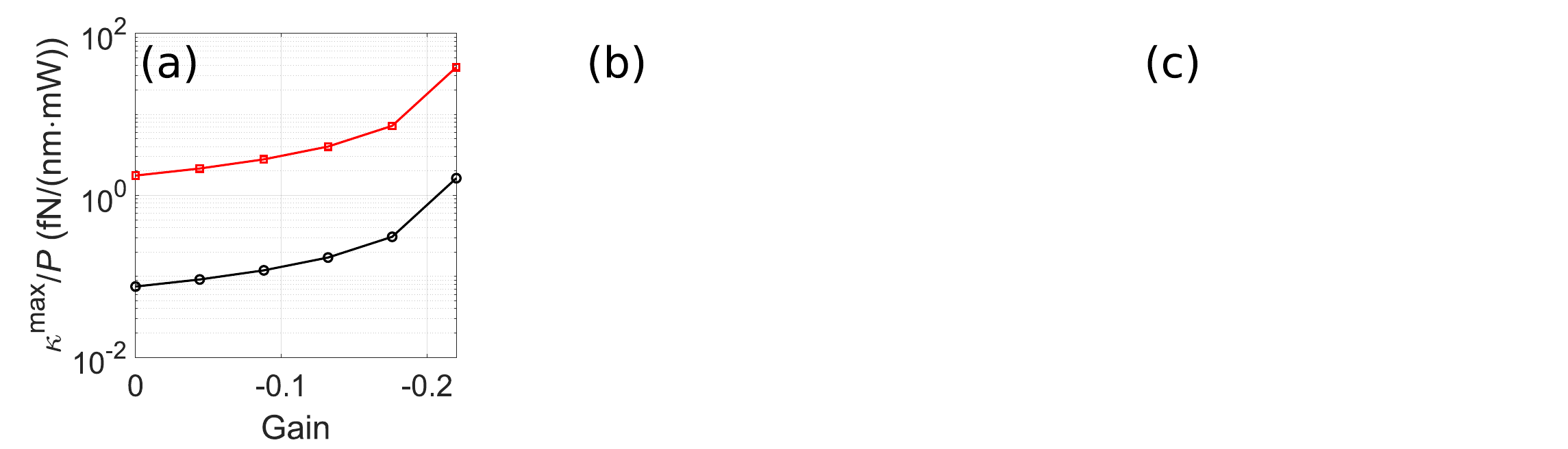}
\caption{In (a), the maximum value of the trapping constants $\kappa_\rho^{\mathrm{max}}$ (black
line) and $\kappa_z^{\mathrm{max}}$ (red line) in the counterpropagating configuration is plotted
as a function of the gain in semi-logarithmic scale. $\kappa_\rho^{\mathrm{max}}$ and
$\kappa_z^{\mathrm{max}}$ are intended as the highest positive values that $\kappa_\rho$ and
$\kappa_z$, respectively, take spanning through the wavelengths according to the trends pictured in
Fig.~\ref{figure2}. Intensity maps of the gradient force in $yz$ plane are shown for $G = 0$ and $G
= - 0.22$, respectively in (b) and (c). The wavelength $\lambda=531.9$ nm is chosen so that the
stiffness obtained in the harmonic approximation assumes the highest values for each of the two
gains. The power of the single Gaussian beam laser is 50 mW.} \label{figure3}
\end{figure}

Figure~\ref{figure3}a shows the maximum value of the trap stiffness as a function of gain. The
onset of a nonlinear behaviour is clearly visible. Both $\kappa_\rho^{\mathrm{max}}$ (black line)
and $\kappa_z^{\mathrm{max}}$ (red line) increase similarly in the logarithmic scale for increasing
gain. While the spring constants give us an idea of the strength of gradient forces for small
displacements around the equilibrium point, to understand how the optical force is spatially
distributed for the whole counter-propagating pattern, Eqs.~\ref{eq:forceDA_CP} must be considered.
This is shown in Figs.~\ref{figure3}b, \ref{figure3}c that represent such distributions in $yz$
plane, respectively at $G = 0$ and $G = - 0.22$. The wavelength $\lambda=531.9$ nm has been chosen
so that the force constant calculated for a small displacements (harmonic) approximation is the
highest for each of the two gains we have considered. The standing wave modulation of the light
intensity (Fig.~\ref{figure1}c) yields the modulated spatial profile of the force. Since we
considered a wavelength on the red-side of the resonance, a nanoshell will be attracted towards the
maximum intensity points, represented with the hottest colors in Fig.~\ref{figure1}c and
corresponding to the minima, blue regions, of the gradient force in Figs.~\ref{figure3}b,
\ref{figure3}c. We note how the gradient force for $G = - 0.22$ (Fig.~\ref{figure3}c) is strikingly
higher than the force for  $G = 0$ (Fig.~\ref{figure3}b), ($i.e.$ $\langle F
\rangle_{\mathrm{DA}}^{\mathrm{max}}(G = - 0.22) / \langle F
\rangle_{\mathrm{DA}}^{\mathrm{max}}(G=0) \sim 2\cdot 10^3$, meaning that, by diverting energy from
the pump, the interplay between the plasmonic resonance and the gain elements is able to produce a
trapping force about three orders of magnitude stronger than the one found in the unpumped system.

\paragraph{Optomechanical position locking and gain$-$assisted channeling.}
To realistically simulate the dynamics of the silver-silica dye-doped nanoshell in the double-beam
OT, we performed Brownian dynamics simulations\cite{volpe2013simulation} in water for different
gain values. Interesting behaviours can arise for the motion of the nanoshells, including
optomechanical position locking and channeling.

The Brownian dynamics takes into account the thermal noise contribution, which tends to  jiggle the
nanoshell in the optical trap. The motion of the particle is, therefore, the result of the
interplay between this random motion and the deterministic optical forces. The time scale on which
the gradient force acts is given by the ratio $\tau_{\mathrm{ot}} = \gamma/\kappa$, where $\gamma$
is the particle friction coefficient in the surrounding fluid, determined by Stokes$'$ law
\cite{jones2015optical,volpe2013simulation}. In our case, $\tau_{\mathrm{ot}}$ is always
significantly greater than the momentum relaxation time $\tau_{\mathrm{in}} = m/\gamma $, so that
inertial effects can be safely neglected. Thus, we can describe the particle Brownian dynamics
through three independent overdamped Langevin equations
\cite{langevin1908theorie,jones2015optical}:
\begin{equation}\label{eq:langevin}
\left\{\begin{array}{ccc} \displaystyle \frac{dx(t)}{dt} & = & \displaystyle
\frac{F_{\mathrm{DA},x}(\mathbf{r},t)}{\gamma} + \sqrt{2D} \, W_x(t) \\ [12pt] \displaystyle
\frac{dy(t)}{dt} & = & \displaystyle \frac{F_{\mathrm{DA},y}(\mathbf{r},t)}{\gamma} + \sqrt{2D} \,
W_y(t) \\ [12pt] \displaystyle \frac{dz(t)}{dt} & = & \displaystyle
\frac{F_{\mathrm{DA},z}(\mathbf{r},t)}{\gamma} + \sqrt{2D} \, W_z(t)
\end{array}\right.
\end{equation}
where $F_{\mathrm{DA},i}$, with $i = x,y,z$, is the $i$-th Cartesian component of the time
dependent optical force in dipole approximation, $D = k_\mathrm{B}T/\gamma$ is  the diffusion
coefficient according to fluctuation-dissipation theorem with $T$ the temperature and
$k_\mathrm{B}$ the Boltzmann's constant, $W_x(t)$, $W_y(t)$ and $W_z(t)$ are the independent white
noises related to the thermal fluctuations. Approximating this ordinary differential equations with
finite difference equations \cite{volpe2013simulation,kloeden1999numerical,callegari2019numerical},
the corresponding system to Eq.~\ref{eq:langevin} is:
\begin{equation}\label{eq:langevin_solve}
\left\{\begin{array}{ccc} \displaystyle x_n & = & \displaystyle x_{n-1} - \frac{\langle F
\rangle_{\mathrm{DA},x,n}}{\gamma} \Delta t + \sqrt{2D\Delta t} \, w_{x,n} \\ [12pt] \displaystyle
y_n & = & \displaystyle y_{n-1} - \frac{\langle F \rangle_{\mathrm{DA},y,n}}{\gamma} \Delta t +
\sqrt{2D\Delta t} \, w_{y,n} \\ [12pt] \displaystyle z_n & = & \displaystyle z_{n-1} -
\frac{\langle F \rangle_{\mathrm{DA},z,n}}{\gamma} \Delta t + \sqrt{2D\Delta t} \, w_{z,n}
\end{array}\right.
\end{equation}
where [$x_n$,$y_n$,$z_n$] represent the position of the particle at time $t_n$, $w_{n,i}$ are the
independent Gaussian random numbers with zero mean and unitary variance that emulate the white
noise, and $n = 1, ..., N$ is an index where $N = 10^6$ is the number of computation steps.
Moreover, $\langle F \rangle_{\mathrm{DA},i,n}$ is the $i$-th component of the optical force in
dipole approximation (Eqs.~\ref{eq:forceDA_CP}) calculated at the $n$-th position. In our
simulation, we consider $T = 300$ K and different time steps, $\Delta t$, depending on the laser
power, {\it e.g.,} in Fig.~\ref{figure1}c a typical value of $\Delta t=10^{-7}$ s has been used for
$P=50$ mW. The sampling time, $t_\mathrm{samp}$, is chosen so that
$t_\mathrm{samp}\gg\tau_{\mathrm{ot}}$. In fact, because of the trap asymmetry we have two
timescales, $\tau_{\rm ot,z}$ and $\tau_{\rm ot,\rho}$, associated to the axial and radial
stiffnesses, respectively, and, $e.g.$, at $P=50$ mW we obtain $\tau_{\rm ot,z}(G=0)\simeq 4\cdot
10^{-6}$ s and $\tau_{\rm ot,\rho}(G=0)\simeq 10^{-4}$ s (for the radial dynamics see Supplementary
Information). Thus, both timescales are much smaller than the sampling time of $t_\mathrm{samp} =
0.1$ s.

\begin{figure}
\centering
\includegraphics[width=\textwidth]{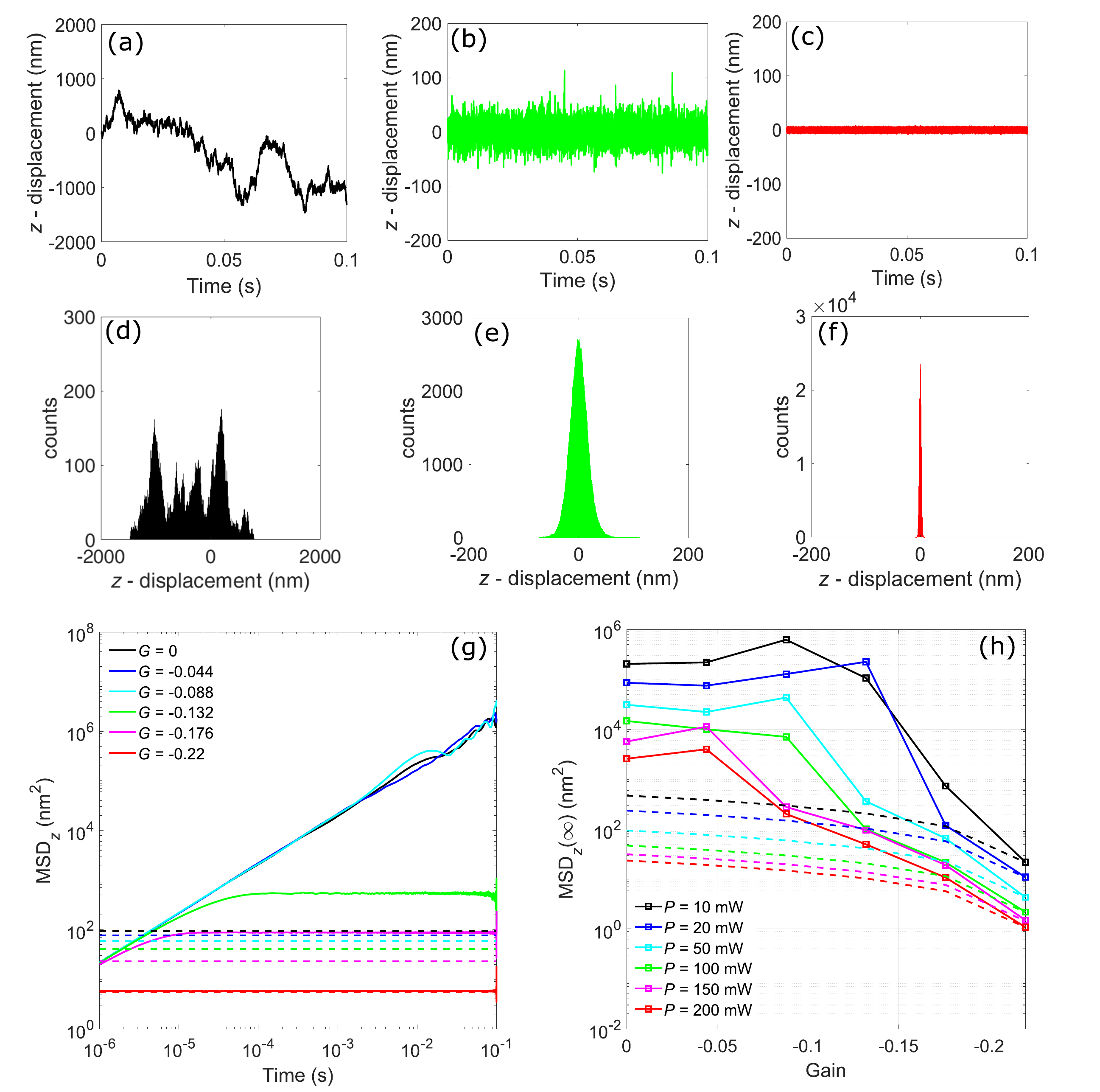}
\caption{(a-c) Brownian dynamics simulation of the dye-doped nanoshell trajectory in water for
gains, $G = 0$ (a), $G = - 0.132$ (b) and $G = - 0.22$ (c) and laser power 50 mW. (d-f) Histograms
of the nanoshell Brownian motion for the position tracks considered in (a-c). In (g), mean square
displacement (MSD) along the $z$ direction for various gains (solid lines). These trends are
compared respectively with the analytic limit of the $\mathrm{MSD}_z$ to infinity, such as
$\mathrm{MSD}_z(\infty) \rightarrow 2k_{\mathrm{B}}T/\kappa_z^{\mathrm{max}}$ (dashed lines). In
(h), the statistical $\mathrm{MSD}_z$ computed at infinity vs gain is shown for different power
values in solid lines. These trends are compared respectively with the analytic limit of the
$\mathrm{MSD}_z$ to infinity, shown in dashed lines. The simulation time is $0.1$ s and the
trapping wavelength has been chosen so that optical trapping forces are maximized for each gain
considered.} \label{figure4}
\end{figure}

With these tools at hand, it is possible to investigate the dynamics of the dye-doped silver-silica
nanoshell. First of all, we consider three different gain $G =0$, $G = - 0.132$ and $G = - 0.22$.
The power in each laser beam is fixed at 50 mW and the trapping wavelength $\lambda$ is chosen so
that the trap stiffnesses are at their maxima for each gain. In Figs.~\ref{figure4}a-c, we show the
simulation of the particle trajectories along the axial direction $z$ that are the starting point
of the analysis (see Supplementary Information for the corresponding movies of the tracking
dynamics projected on $yz$ plane). When $G=0$, Fig. \ref{figure4}a, the particle is not trapped,
the nanoshell jumps in the standing wave maxima and therefore its position is never locked in any
of the optical potential minima exploring more than $2$ $\mu$m within $0.1$ s. When $G = - 0.132$,
Fig.~\ref{figure4}b, the particle trajectory fluctuates around its equilibrium position ($z=0$) and
the nanoshell is trapped in the central intensity maximum of the standing wave. For the highest
gain $G = - 0.22$, (Fig.~\ref{figure4}c), the tracking profile is tightly locked in the $z=0$
position with much smaller fluctuations. We can further quantify our analysis by calculating the
position histograms and the mean square displacements of these trajectories. Fig.~\ref{figure4}d-f
show the position histograms corresponding to the three considered gain. For the case of $G = 0$,
Fig.~\ref{figure4}d, the nanoshell interacts very weakly with the trapping beam and the thermal
noise contribution to the dynamics is much larger than the trapping potential depth. The particle
explores several standing wave intensity maxima that show up in the structure of the histogram as
different peaks with widths of the order of $\sim 200$ nm, of the same order of the standing wave
periodicity. Instead, increasing gain ($G = - 0.132$, Fig.~\ref{figure4}e), the particle position
appears locked to the central intensity maximum of the double-beam OT. The thermal noise
contribution is smaller than the trapping potential depth and the nanoparticle explores only a $19$
nm region around the central high intensity spot. For the highest gain $G = - 0.22$,
(Fig.~\ref{figure4}f), the position distribution is extremely narrow, and the gain-enhanced
gradient force tightly confines the particle within only $2$ nm.

We now consider the behaviour of the particle mean square displacement, ${\mathrm {MSD}}_z$, along
axial $z$ direction as a function of gain (See the Supplementary Information for the analysis of
MSD along the transverse direction, $\rho$). This quantity compute{\color{red}s} the deviation of
the particle position (Eqs.~\ref{eq:langevin_solve}), $z_n$, from its preceding position for each
time interval $\Delta t$, $\mathrm{MSD}_z (t_n)=\langle|z_n - z_{n-m}|^2\rangle_m=\sum_{m=1}^n|z_n
- z_{n-m}|^2/n$. Therefore, in Fig.~\ref{figure4}g, we have considered the calculated statistical
${\mathrm {MSD}}_z$ versus time for increasing the gain (solid lines) and compared them to the MSD
analytically calculated considering the harmonic approximation for small displacements around $z=0$
(dashed lines) where ${\mathrm {MSD}}_z$ can be calculated analytically in terms of trap
stiffnesses\cite{polimeno2018optical,jones2015optical}, ${\mathrm{MSD}}_z(t) = \left( 2 k_{\mathrm
B} T/\kappa_z^{\mathrm{max}} \right) \left[ 1 - \exp\left(- t / \tau_{\mathrm{to}}\right) \right]$.
Hence, at long times we have that ${\mathrm{MSD}}_z(\infty) \rightarrow
2k_{\mathrm{B}}T/\kappa_z^{\mathrm{max}}$. We note that in Fig.~\ref{figure4}g the value of
${\mathrm{MSD}}_z$ approaches to ${\mathrm{MSD}}_z(\infty)$ only for the highest gain ($G = -
0.22$). This occurs when the potential energy barrier is so high that the thermal fluctuations do
not have enough energy to drive the hopping between the different potential wells in the standing
wave. Consequently, the small displacements approximation gives a reliable description of the
dynamics and the theoretical trapping constants (Fig.~\ref{figure2}i) can be safely used to predict
the MSD values. On the other hand, for the lower gains the ${\mathrm{MSD}}_z$ from the simulation
is larger than the values predicted by the small displacements approximation with increasing
discrepancies as the gain decreases. At zero gain optical potential barriers are much lower than
the energy of the thermal fluctuations. Thus, the particle hops between several standing wave
maxima and the small displacement approximation does not give a reliable description of the
simulated particle dynamics.

Another parameter of the double-beam OT is the laser beam power that will crucially determine the
optical trapping dynamics of the dye-doped nanoshell. In Fig.~\ref{figure4}h (solid lines) we study
the MSD at infinity, ${\mathrm{MSD}}_z(\infty)$, as a function of gain parameterized for different
power $P$. Since optical forces increase with laser power strengthening the trap, for each gain
value ${\mathrm{MSD}}_z(\infty)$ decreases with increasing power and the tightest confinement
occurs at high gain values and high power. Also in this case, we compare the analysis of the
Brownian dynamics results with the analytical values of ${\mathrm{MSD}}_z(\infty)$ obtained using
the harmonic approximation for the standing wave trapping (dashed lines). For each power, the
simulation results tend to overlap with the harmonic approximation as gain increases. The trapping
of the nanoshell is so efficient at high gain that the dynamics can be faithfully studied in the
single-trap harmonic approximation.

Finally, we describe simulations of a possible practical situation when dye-doped nanoshell
particles are immersed in a microfluidic flow and laser light is selectively tuned across the
gain-enhanced resonance for position locking, channeling or sorting in a fashion similar to what
has been developed for cold atoms\cite{salomon1987channeling,meschede2003atomic}. We consider
nanoshell particles in water with a high gain $G = - 0.22$ and tune the double-beam OT wavelength
within a fraction of nanometer across the sharp resonance peak (see Fig.~\ref{figure2}g) so that
the sign of the gradient force is switched (see Figs.~\ref{figure2}h,i). In the simulations the
power of the single Gaussian beam is fixed at $50$ mW and the trajectory is simulated for $0.1$ s.
In order to emulate a generic particle flux, we have considered three identical nanoshells whose
dynamics evolve from three different starting positions $\mathbf{r}_{01} = (0,-400,-250)$ nm,
$\mathbf{r}_{02} = (0,- 400,0)$ nm, $\mathbf{r}_{03} = (0,-400,250)$ nm with constant flow velocity
$\mathbf{v}_0 =$ (0,1,0) $\mathrm{mm}/\mathrm{s}$. The reference system origin is taken in
correspondence of the center of the double-beam OT.
\begin{figure}
\centering
\includegraphics[width=\textwidth]{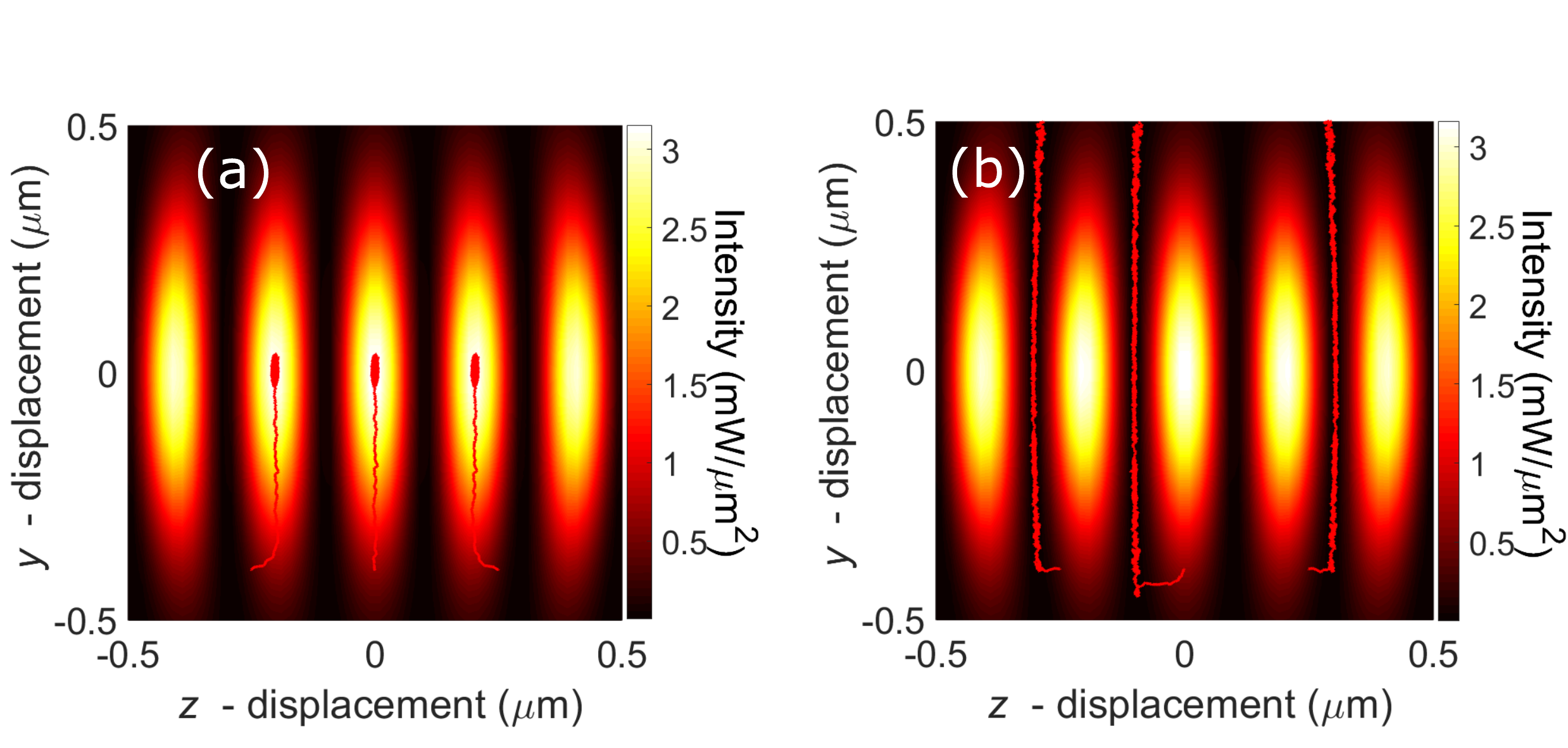}
\caption{Nanoparticles trajectory projections on $yz$ plane are plotted for $G = - 0.22$ both for
$\lambda \simeq 531.9$ nm (a), when $\mathrm{Re}\{\alpha\}/\varepsilon_0 \simeq 5.02\cdot10^6$
nm$^3$, and for $\lambda \simeq 531.5$ nm (b), when $\mathrm{Re}\{\alpha\}/\varepsilon_0 \simeq
-5.88\cdot10^6$ {nm}$^3$. In (a), the radiation intensity maxima correspond to the minima of the
optical potential. Therefore, these are attraction points for nanoshells that flow into the
standing wave region from three different points $\mathbf{r}_{01} = (0,-400,-250)$ nm,
$\mathbf{r}_{02} =$ (0,-400,0) nm, $\mathbf{r}_{03} = (0,-400,250)$ nm with constant flow velocity
$\mathbf{v}_0 =$ (0,1,0) $\mathrm{mm}/\mathrm{s}$. Instead, in (b), the intensity maxima correspond
to the maximum of the optical potential and the gradient force is maximally repulsive. In this way,
the motion of the three nanoshells (with the same initial conditions of the previous situation) is
channeled. Each trajectory is simulated for 0.1 s and the power of the single Gaussian beam is
fixed at 50 mW.} \label{figure5}
\end{figure}
First, we simulate the particle motion with a double-beam OT wavelength of $\lambda_{\mathrm{max}}
\simeq 531.9$ nm, corresponding to the maximum trapping force. Due to the positive sign of the
optical force components, the flowing nanoshells are {\it high-field seekers} and their position
evolve towards specific equilibrium points within the standing wave maxima. Figure~\ref{figure5}a
show the trajectories, projected on the $yz$ plane, of the three particles superposed with the
standing wave intensity pattern. Conversely, when the wavelength is tuned to
$\lambda_{\mathrm{min}} \simeq 531.5$ nm, the optical force is reversed and the flowing nanoshells
are {\it low-field seekers}. Figure~\ref{figure5}b shows that the standing wave maxima act as
repulsive spots and the particles are efficiently channelled through the standing wave intensity
minima. Movies for the two situations, showing position locking and gain-assisted channeling, are
shown as Supplementary Information.

We note that the role of radiation pressure by the pump field has not been considered in our
analysis. Although it can be a source of instability for optical trapping, a double beam
configuration can be always applied so that also for the pump field scattering forces are balanced
in the trap region.

\section{Conclusion} In conclusion, we described a systematic investigation of the behaviours of hybrid core-shell
nanoparticles with gain in a double-beam optical trap. We highlighted the non-linear optical
scaling of optical forces. Furthermore, we performed Brownian dynamics simulations in water for
realistic situations, where we directly observe how the particle dynamics is more confined for
increasing gain and power. Finally, we showed that by tuning the light wavelength within a fraction
of a nanometer it is possible to switch the sign of the optical forces and use the dual-beam
configuration in a micro-fluidic flow for particle position locking, when the wavelength is
red-detuned with respect to the gain-enhanced resonance, or gain-assisted channelling, when the
wavelength is blue-detuned with respect to the resonance. Our results are obtained in dipole
approximation for a small nanoshell. For larger, non-spherical, or aggregated particles, a T-matrix
approach can be used \cite{pezzi2019resonant}. However, we expect the general features of the
nonlinear gain-assisted optomechanical effects described above to still hold. This work opens
perspectives for highly controlled optical sorting and guiding of hybrid particles in
microfluidics.


\paragraph{Supporting Information.} Gain$-$assisted optical trapping in single-beam optical tweezers, Brownian dynamics simulations in the transverse radial direction, supplementary movies.

\begin{acknowledgement}
We acknowledge financial contribution from the agreement ASI-INAF n.2018-16-HH.0, project ``SPACE
Tweezers" and from USFQ's 2019 Poligrants Program.
\end{acknowledgement}




\bibliographystyle{achemso}

\providecommand{\latin}[1]{#1}
\makeatletter
\providecommand{\doi}
  {\begingroup\let\do\@makeother\dospecials
  \catcode`\{=1 \catcode`\}=2 \doi@aux}
\providecommand{\doi@aux}[1]{\endgroup\texttt{#1}}
\makeatother
\providecommand*\mcitethebibliography{\thebibliography}
\csname @ifundefined\endcsname{endmcitethebibliography}
  {\let\endmcitethebibliography\endthebibliography}{}


\begin{mcitethebibliography}{41}
\providecommand*\natexlab[1]{#1}
\providecommand*\mciteSetBstSublistMode[1]{}
\providecommand*\mciteSetBstMaxWidthForm[2]{}
\providecommand*\mciteBstWouldAddEndPuncttrue
  {\def\EndOfBibitem{\unskip.}}
\providecommand*\mciteBstWouldAddEndPunctfalse
  {\let\EndOfBibitem\relax}
\providecommand*\mciteSetBstMidEndSepPunct[3]{}
\providecommand*\mciteSetBstSublistLabelBeginEnd[3]{}
\providecommand*\EndOfBibitem{}
\mciteSetBstSublistMode{f}
\mciteSetBstMaxWidthForm{subitem}{(\alph{mcitesubitemcount})}
\mciteSetBstSublistLabelBeginEnd
  {\mcitemaxwidthsubitemform\space}
  {\relax}
  {\relax}

\bibitem[Ashkin \latin{et~al.}(1986)Ashkin, Dziedzic, Bjorkholm, and
  Chu]{Ashkin1986}
Ashkin,~A.; Dziedzic,~J.; Bjorkholm,~J.; Chu,~S. Observation of a Single-Beam
  Gradient Optical Trap for Dielectric Particles. \emph{Opt. Lett.}
  \textbf{1986}, \emph{11}, 288--290\relax
\mciteBstWouldAddEndPuncttrue
\mciteSetBstMidEndSepPunct{\mcitedefaultmidpunct}
{\mcitedefaultendpunct}{\mcitedefaultseppunct}\relax
\EndOfBibitem
\bibitem[Jones \latin{et~al.}(2015)Jones, Marag{\`o}, and
  Volpe]{jones2015optical}
Jones,~P.~H.; Marag{\`o},~O.~M.; Volpe,~G. \emph{Optical Tweezers: Principles
  and Applications}; Cambridge University Press: Cambridge, UK, 2015\relax
\mciteBstWouldAddEndPuncttrue
\mciteSetBstMidEndSepPunct{\mcitedefaultmidpunct}
{\mcitedefaultendpunct}{\mcitedefaultseppunct}\relax
\EndOfBibitem
\bibitem[Jon{\'a}\v{s} and Zem{\'a}nek(2008)Jon{\'a}\v{s}, and
  Zem{\'a}nek]{JonasEPH08}
Jon{\'a}\v{s},~A.; Zem{\'a}nek,~P. {Light at Work: {T}he Use of Optical Forces
  for Particle Manipulation, Sorting, and Analysis.} \emph{Electophoresis}
  \textbf{2008}, \emph{29}, 4813--4851\relax
\mciteBstWouldAddEndPuncttrue
\mciteSetBstMidEndSepPunct{\mcitedefaultmidpunct}
{\mcitedefaultendpunct}{\mcitedefaultseppunct}\relax
\EndOfBibitem
\bibitem[Polimeno \latin{et~al.}(2018)Polimeno, Magazz\`{u}, Iat\`{\i}, Patti,
  Saija, Degli Esposti~Boschi, Donato, Gucciardi, Jones, Volpe, and
  Marag\`{o}]{polimeno2018optical}
Polimeno,~P.; Magazz\`{u},~A.; Iat\`{\i},~M.~A.; Patti,~F.; Saija,~R.; Degli
  Esposti~Boschi,~C.; Donato,~M.~G.; Gucciardi,~P.~G.; Jones,~P.~H.; Volpe,~G.;
  Marag\`{o},~O.~M. Optical Tweezers and their Applications. \emph{J. Quant.
  Spectrosc. Radiat. Transf.} \textbf{2018}, \emph{218}, 131--150\relax
\mciteBstWouldAddEndPuncttrue
\mciteSetBstMidEndSepPunct{\mcitedefaultmidpunct}
{\mcitedefaultendpunct}{\mcitedefaultseppunct}\relax
\EndOfBibitem
\bibitem[Marag\`{o} \latin{et~al.}(2013)Marag\`{o}, Jones, Gucciardi, Volpe,
  and Ferrari]{Marago2013}
Marag\`{o},~O.~M.; Jones,~P.~H.; Gucciardi,~P.~G.; Volpe,~G.; Ferrari,~A.~C.
  Optical Trapping and Manipulation of Nanostructures. \emph{Nat. Nanotechnol.}
  \textbf{2013}, \emph{8}, 807--819\relax
\mciteBstWouldAddEndPuncttrue
\mciteSetBstMidEndSepPunct{\mcitedefaultmidpunct}
{\mcitedefaultendpunct}{\mcitedefaultseppunct}\relax
\EndOfBibitem
\bibitem[Spesyvtseva and Dholakia(2016)Spesyvtseva, and
  Dholakia]{Spesyvtseva2016}
Spesyvtseva,~S. E.~S.; Dholakia,~K. Trapping in a Material World. \emph{ACS
  Photonics} \textbf{2016}, \emph{3}, 719--736\relax
\mciteBstWouldAddEndPuncttrue
\mciteSetBstMidEndSepPunct{\mcitedefaultmidpunct}
{\mcitedefaultendpunct}{\mcitedefaultseppunct}\relax
\EndOfBibitem
\bibitem[Yan \latin{et~al.}(2013)Yan, Pelton, Vigderman, Zubarev, and
  Scherer]{YanACSNANO13b}
Yan,~Z.; Pelton,~M.; Vigderman,~L.; Zubarev,~E.; Scherer,~N. Why Single-Beam
  Optical Tweezers Trap Gold Nanowires in Three Dimensions. \emph{ACS Nano}
  \textbf{2013}, \emph{7}, 8794--8800\relax
\mciteBstWouldAddEndPuncttrue
\mciteSetBstMidEndSepPunct{\mcitedefaultmidpunct}
{\mcitedefaultendpunct}{\mcitedefaultseppunct}\relax
\EndOfBibitem
\bibitem[Irrera \latin{et~al.}(2016)Irrera, Magazz{\`u}, Artoni, Simpson,
  Hanna, Jones, Priolo, Gucciardi, and Marag{\`o}]{Irrera2016}
Irrera,~A.; Magazz{\`u},~A.; Artoni,~P.; Simpson,~S.~H.; Hanna,~S.;
  Jones,~P.~H.; Priolo,~F.; Gucciardi,~P.~G.; Marag{\`o},~O.~M. Photonic Torque
  Microscopy of the Nonconservative Force Field for Optically Trapped Silicon
  Nanowires. \emph{Nano Lett.} \textbf{2016}, \emph{16}, 4181--4188\relax
\mciteBstWouldAddEndPuncttrue
\mciteSetBstMidEndSepPunct{\mcitedefaultmidpunct}
{\mcitedefaultendpunct}{\mcitedefaultseppunct}\relax
\EndOfBibitem
\bibitem[Van~der Horst \latin{et~al.}(2008)Van~der Horst, van Oostrum, Moroz,
  van Blaaderen, and Dogterom]{VanderHorst2008}
Van~der Horst,~A.; van Oostrum,~P.~D.; Moroz,~A.; van Blaaderen,~A.;
  Dogterom,~M. High Trapping Forces for High-Refractive Index Particles Trapped
  in Dynamic Arrays of Counterpropagating Optical Tweezers. \emph{Appl. Opt.}
  \textbf{2008}, \emph{47}, 3196--3202\relax
\mciteBstWouldAddEndPuncttrue
\mciteSetBstMidEndSepPunct{\mcitedefaultmidpunct}
{\mcitedefaultendpunct}{\mcitedefaultseppunct}\relax
\EndOfBibitem
\bibitem[Marag\`{o} \latin{et~al.}(2008)Marag\`{o}, Jones, Bonaccorso,
  Scardaci, Gucciardi, Rozhin, and Ferrari]{Marago2008a}
Marag\`{o},~O.~M.; Jones,~P.~H.; Bonaccorso,~F.; Scardaci,~V.;
  Gucciardi,~P.~G.; Rozhin,~A.~G.; Ferrari,~A.~C. Femtonewton Force Sensing
  with Optically Trapped Nanotubes. \emph{Nano Lett.} \textbf{2008}, \emph{8},
  3211--3216\relax
\mciteBstWouldAddEndPuncttrue
\mciteSetBstMidEndSepPunct{\mcitedefaultmidpunct}
{\mcitedefaultendpunct}{\mcitedefaultseppunct}\relax
\EndOfBibitem
\bibitem[Donato \latin{et~al.}(2018)Donato, Messina, Foti, Smart, Jones,
  Iat\`{\i}, Saija, Gucciardi, and Marag\`{o}]{Donato2018}
Donato,~M.~G.; Messina,~E.; Foti,~A.; Smart,~T.~J.; Jones,~P.~H.;
  Iat\`{\i},~M.~A.; Saija,~R.; Gucciardi,~P.~G.; Marag\`{o},~O.~M. Optical
  Trapping and Optical Force Positioning of Two-Dimensional Materials.
  \emph{Nanoscale} \textbf{2018}, \emph{10}, 1245--1255\relax
\mciteBstWouldAddEndPuncttrue
\mciteSetBstMidEndSepPunct{\mcitedefaultmidpunct}
{\mcitedefaultendpunct}{\mcitedefaultseppunct}\relax
\EndOfBibitem
\bibitem[Spadaro \latin{et~al.}(2015)Spadaro, Iat\`{\i}, Donato, Gucciardi,
  Saija, Cherlakola, Scaramuzza, Amendola, and Marag\`{o}]{Spadaro2015}
Spadaro,~D.; Iat\`{\i},~M.~A.; Donato,~M.~G.; Gucciardi,~P.~G.; Saija,~R.;
  Cherlakola,~A.~R.; Scaramuzza,~S.; Amendola,~V.; Marag\`{o},~O.~M. Scaling of
  Optical Forces on Au--PEG Core--Shell Nanoparticles. \emph{RSC Adv.}
  \textbf{2015}, \emph{5}, 93139--93146\relax
\mciteBstWouldAddEndPuncttrue
\mciteSetBstMidEndSepPunct{\mcitedefaultmidpunct}
{\mcitedefaultendpunct}{\mcitedefaultseppunct}\relax
\EndOfBibitem
\bibitem[Spadaro \latin{et~al.}(2016)Spadaro, Iat\`{\i}, P{\'e}rez-Pi{\~n}eiro,
  V{\'a}zquez-V{\'a}zquez, Correa-Duarte, Donato, Gucciardi, Saija, Strangi,
  and Marag\`{o}]{Spadaro2016}
Spadaro,~D.; Iat\`{\i},~M.~A.; P{\'e}rez-Pi{\~n}eiro,~J.;
  V{\'a}zquez-V{\'a}zquez,~C.; Correa-Duarte,~M.~A.; Donato,~M.~G.;
  Gucciardi,~P.~G.; Saija,~R.; Strangi,~G.; Marag\`{o},~O.~M. Optical Trapping
  of Plasmonic Mesocapsules: Enhanced Optical Forces and SERS. \emph{J. Phys.
  Chem. C} \textbf{2016}, \emph{121}, 691--700\relax
\mciteBstWouldAddEndPuncttrue
\mciteSetBstMidEndSepPunct{\mcitedefaultmidpunct}
{\mcitedefaultendpunct}{\mcitedefaultseppunct}\relax
\EndOfBibitem
\bibitem[Lehmuskero \latin{et~al.}(2015)Lehmuskero, Johansson,
  Rubinsztein-Dunlop, Tong, and Kall]{Lehmuskero2015}
Lehmuskero,~A.; Johansson,~P.; Rubinsztein-Dunlop,~H.; Tong,~L.; Kall,~M. Laser
  Trapping of Colloidal Metal Nanoparticles. \emph{ACS Nano} \textbf{2015},
  \emph{9}, 3453--3469\relax
\mciteBstWouldAddEndPuncttrue
\mciteSetBstMidEndSepPunct{\mcitedefaultmidpunct}
{\mcitedefaultendpunct}{\mcitedefaultseppunct}\relax
\EndOfBibitem
\bibitem[Brzobohat\'{y} \latin{et~al.}(2015)Brzobohat\'{y}, \v{S}iler, Trojek,
  Chv\'{a}tal, Kar\'{a}sek, Pat\'{a}k, Pokorn\'{a}, Mika, and
  Zem\'{a}nek]{BrzobohatySR15}
Brzobohat\'{y},~O.; \v{S}iler,~M.; Trojek,~J.; Chv\'{a}tal,~L.;
  Kar\'{a}sek,~V.; Pat\'{a}k,~A.; Pokorn\'{a},~Z.; Mika,~F.; Zem\'{a}nek,~P.
  Three-Dimensional Optical Trapping of a Plasmonic Nanoparticle using Low
  Numerical Aperture Optical Tweezers. \emph{Sci. Rep.} \textbf{2015},
  \emph{5}, 8106\relax
\mciteBstWouldAddEndPuncttrue
\mciteSetBstMidEndSepPunct{\mcitedefaultmidpunct}
{\mcitedefaultendpunct}{\mcitedefaultseppunct}\relax
\EndOfBibitem
\bibitem[Messina \latin{et~al.}(2015)Messina, Donato, Zimbone, Saija,
  Iat\`{\i}, Calcagno, Fragala, Compagnini, D'Andrea, Foti, Gucciardi, and
  Marag\`{o}]{Messina2015}
Messina,~E.; Donato,~M.~G.; Zimbone,~M.; Saija,~R.; Iat\`{\i},~M.~A.;
  Calcagno,~L.; Fragala,~M.~E.; Compagnini,~G.; D'Andrea,~C.; Foti,~A.;
  Gucciardi,~P.~G.; Marag\`{o},~O.~M. Optical Trapping of Silver Nanoplatelets.
  \emph{Opt. Express} \textbf{2015}, \emph{23}, 8720--8730\relax
\mciteBstWouldAddEndPuncttrue
\mciteSetBstMidEndSepPunct{\mcitedefaultmidpunct}
{\mcitedefaultendpunct}{\mcitedefaultseppunct}\relax
\EndOfBibitem
\bibitem[Ashkin(1970)]{Ashkin1970}
Ashkin,~A. Acceleration and Trapping of Particles by Radiation Pressure.
  \emph{Phys. Rev. Lett.} \textbf{1970}, \emph{24}, 156\relax
\mciteBstWouldAddEndPuncttrue
\mciteSetBstMidEndSepPunct{\mcitedefaultmidpunct}
{\mcitedefaultendpunct}{\mcitedefaultseppunct}\relax
\EndOfBibitem
\bibitem[Zem{\'a}nek \latin{et~al.}(2002)Zem{\'a}nek, Jon{\'a}\v{s}, and
  Li\v{s}ka]{ZemanekJOSAA02}
Zem{\'a}nek,~P.; Jon{\'a}\v{s},~A.; Li\v{s}ka,~M. {Simplified Description of
  Optical Forces Acting on a Nanoparticle in the {G}aussian Standing Wave}.
  \emph{J. Opt. Soc. Am. A} \textbf{2002}, \emph{19}, 1025--1034\relax
\mciteBstWouldAddEndPuncttrue
\mciteSetBstMidEndSepPunct{\mcitedefaultmidpunct}
{\mcitedefaultendpunct}{\mcitedefaultseppunct}\relax
\EndOfBibitem
\bibitem[\v{C}i\v{z}m{\'a}r \latin{et~al.}(2006)\v{C}i\v{z}m{\'a}r, \v{S}iler,
  and Zem{\'a}nek]{CizmarAPB06}
\v{C}i\v{z}m{\'a}r,~T.; \v{S}iler,~M.; Zem{\'a}nek,~P. {An Optical Nanotrap
  Array Movable over a Milimetre Range}. \emph{Appl. Phys. B} \textbf{2006},
  \emph{84}, 197--203\relax
\mciteBstWouldAddEndPuncttrue
\mciteSetBstMidEndSepPunct{\mcitedefaultmidpunct}
{\mcitedefaultendpunct}{\mcitedefaultseppunct}\relax
\EndOfBibitem
\bibitem[Gherardi \latin{et~al.}(2008)Gherardi, Carruthers, \v{C}i\v{z}m{\'a}r,
  Wright, and Dholakia]{GherardiAPL08}
Gherardi,~D.~M.; Carruthers,~A.~E.; \v{C}i\v{z}m{\'a}r,~T.; Wright,~E.~M.;
  Dholakia,~K. {A Dual Beam Photonic Crystal Fibre Trap for Microscopic
  Particles}. \emph{Appl. Phys. Lett.} \textbf{2008}, \emph{93}, 041110\relax
\mciteBstWouldAddEndPuncttrue
\mciteSetBstMidEndSepPunct{\mcitedefaultmidpunct}
{\mcitedefaultendpunct}{\mcitedefaultseppunct}\relax
\EndOfBibitem
\bibitem[Singer \latin{et~al.}(2003)Singer, Frick, Bernet, and
  Ritsch-Marte]{Singer2003}
Singer,~W.; Frick,~M.; Bernet,~S.; Ritsch-Marte,~M. Self-Organized Array of
  Regularly Spaced Microbeads in a Fiber-Optical Trap. \emph{J. Opt. Soc. Am.
  B} \textbf{2003}, \emph{20}, 1568--1574\relax
\mciteBstWouldAddEndPuncttrue
\mciteSetBstMidEndSepPunct{\mcitedefaultmidpunct}
{\mcitedefaultendpunct}{\mcitedefaultseppunct}\relax
\EndOfBibitem
\bibitem[Donato \latin{et~al.}(2019)Donato, Brzobohaty, Simpson, Irrera,
  Leonardi, Lo~Faro, Svak, Marago, and Zem{\'a}nek]{donato2019optical}
Donato,~M.~G.; Brzobohaty,~O.; Simpson,~S.~H.; Irrera,~A.; Leonardi,~A.~A.;
  Lo~Faro,~M.~J.; Svak,~V.; Marago,~O.~M.; Zem{\'a}nek,~P. Optical Trapping,
  Optical Binding, and Rotational Dynamics of Silicon Nanowires in
  Counter-Propagating Beams. \emph{Nano Letters} \textbf{2019}, \emph{19},
  342--352\relax
\mciteBstWouldAddEndPuncttrue
\mciteSetBstMidEndSepPunct{\mcitedefaultmidpunct}
{\mcitedefaultendpunct}{\mcitedefaultseppunct}\relax
\EndOfBibitem
\bibitem[Veltri and Aradian(2012)Veltri, and Aradian]{veltri2012optical}
Veltri,~A.; Aradian,~A. Optical Response of a Metallic Nanoparticle Immersed in
  a Medium with Optical Gain. \emph{Physical Review B} \textbf{2012},
  \emph{85}, 115429\relax
\mciteBstWouldAddEndPuncttrue
\mciteSetBstMidEndSepPunct{\mcitedefaultmidpunct}
{\mcitedefaultendpunct}{\mcitedefaultseppunct}\relax
\EndOfBibitem
\bibitem[Infusino \latin{et~al.}(2014)Infusino, De~Luca, Veltri,
  Vazquez-Vazquez, Correa-Duarte, Dhama, and Strangi]{infusino2014loss}
Infusino,~M.; De~Luca,~A.; Veltri,~A.; Vazquez-Vazquez,~C.;
  Correa-Duarte,~M.~A.; Dhama,~R.; Strangi,~G. Loss-Mitigated Collective
  Resonances in Gain-Assisted Plasmonic Mesocapsules. \emph{ACS Photonics}
  \textbf{2014}, \emph{1}, 371--376\relax
\mciteBstWouldAddEndPuncttrue
\mciteSetBstMidEndSepPunct{\mcitedefaultmidpunct}
{\mcitedefaultendpunct}{\mcitedefaultseppunct}\relax
\EndOfBibitem
\bibitem[Caligiuri \latin{et~al.}(2016)Caligiuri, Pezzi, Veltri, and
  De~Luca]{caligiuri2016resonant}
Caligiuri,~V.; Pezzi,~L.; Veltri,~A.; De~Luca,~A. Resonant Gain Singularities
  in 1d and 3d Metal/Dielectric Multilayered Nanostructures. \emph{ACS nano}
  \textbf{2016}, \emph{11}, 1012--1025\relax
\mciteBstWouldAddEndPuncttrue
\mciteSetBstMidEndSepPunct{\mcitedefaultmidpunct}
{\mcitedefaultendpunct}{\mcitedefaultseppunct}\relax
\EndOfBibitem
\bibitem[Veltri \latin{et~al.}(2016)Veltri, Chipouline, and
  Aradian]{veltri2016multipolar}
Veltri,~A.; Chipouline,~A.; Aradian,~A. Multipolar, Time-Dynamical Model for
  the Loss Compensation and Lasing of a Spherical Plasmonic Nanoparticle Spaser
  Immersed in an Active Gain Medium. \emph{Scientific Reports} \textbf{2016},
  \emph{6}, 33018\relax
\mciteBstWouldAddEndPuncttrue
\mciteSetBstMidEndSepPunct{\mcitedefaultmidpunct}
{\mcitedefaultendpunct}{\mcitedefaultseppunct}\relax
\EndOfBibitem
\bibitem[Pezzi \latin{et~al.}(2019)Pezzi, Iati, Saija, De~Luca, and
  Marago]{pezzi2019resonant}
Pezzi,~L.; Iati,~M.~A.; Saija,~R.; De~Luca,~A.; Marago,~O.~M. Resonant Coupling
  and Gain Singularities in Metal/Dielectric Multishells: Quasi-Static Versus
  T-Matrix Calculations. \emph{The Journal of Physical Chemistry C}
  \textbf{2019}, \emph{123}, 29291--29297\relax
\mciteBstWouldAddEndPuncttrue
\mciteSetBstMidEndSepPunct{\mcitedefaultmidpunct}
{\mcitedefaultendpunct}{\mcitedefaultseppunct}\relax
\EndOfBibitem
\bibitem[Johnson and Christy(1972)Johnson, and Christy]{johnson1972optical}
Johnson,~P.~B.; Christy,~R.-W. Optical Constants of the Noble Metals.
  \emph{Physical Review B} \textbf{1972}, \emph{6}, 4370\relax
\mciteBstWouldAddEndPuncttrue
\mciteSetBstMidEndSepPunct{\mcitedefaultmidpunct}
{\mcitedefaultendpunct}{\mcitedefaultseppunct}\relax
\EndOfBibitem
\bibitem[Chaumet and Nieto-Vesperinas(2000)Chaumet, and
  Nieto-Vesperinas]{chaumet2000time}
Chaumet,~P.; Nieto-Vesperinas,~M. Time-Averaged Total Force on a Dipolar Sphere
  in an Electromagnetic Field. \emph{Opt. Lett.} \textbf{2000}, \emph{25},
  1065--1067\relax
\mciteBstWouldAddEndPuncttrue
\mciteSetBstMidEndSepPunct{\mcitedefaultmidpunct}
{\mcitedefaultendpunct}{\mcitedefaultseppunct}\relax
\EndOfBibitem
\bibitem[Arias-Gonz{\'a}lez and Nieto-Vesperinas(2003)Arias-Gonz{\'a}lez, and
  Nieto-Vesperinas]{arias2003optical}
Arias-Gonz{\'a}lez,~J.~R.; Nieto-Vesperinas,~M. Optical Forces on Small
  Particles: Attractive and Repulsive Nature and Plasmon-Resonance Conditions.
  \emph{J. Opt. Soc. Am. A} \textbf{2003}, \emph{20}, 1201--1209\relax
\mciteBstWouldAddEndPuncttrue
\mciteSetBstMidEndSepPunct{\mcitedefaultmidpunct}
{\mcitedefaultendpunct}{\mcitedefaultseppunct}\relax
\EndOfBibitem
\bibitem[Albaladejo \latin{et~al.}(2009)Albaladejo, Laroche, and
  S{\'a}enz]{albaladejo2009scattering}
Albaladejo,~S.; Laroche,~M. I. M.~M.; S{\'a}enz,~J.~J. Scattering Forces from
  the Curl of the Spin Angular Momentum of a Light Field. \emph{Phys. Rev.
  Lett.} \textbf{2009}, \emph{102}, 113602\relax
\mciteBstWouldAddEndPuncttrue
\mciteSetBstMidEndSepPunct{\mcitedefaultmidpunct}
{\mcitedefaultendpunct}{\mcitedefaultseppunct}\relax
\EndOfBibitem
\bibitem[Gao \latin{et~al.}(2017)Gao, Ding, Nieto-Vesperinas, Ding, Rahman,
  Zhang, Lim, and Qiu]{gao2017optical}
Gao,~D.; Ding,~W.; Nieto-Vesperinas,~M.; Ding,~X.; Rahman,~M.; Zhang,~T.;
  Lim,~C.; Qiu,~C.-W. Optical Manipulation from the Microscale to the
  Nanoscale: Fundamentals, Advances and Prospects. \emph{Light Sci. Appl.}
  \textbf{2017}, \emph{6}, e17039\relax
\mciteBstWouldAddEndPuncttrue
\mciteSetBstMidEndSepPunct{\mcitedefaultmidpunct}
{\mcitedefaultendpunct}{\mcitedefaultseppunct}\relax
\EndOfBibitem
\bibitem[Zem{\'a}nek \latin{et~al.}(1998)Zem{\'a}nek, Jon{\'a}{\v{s}},
  {\v{S}}r{\'a}mek, and Li{\v{s}}ka]{zemanek1998optical}
Zem{\'a}nek,~P.; Jon{\'a}{\v{s}},~A.; {\v{S}}r{\'a}mek,~L.; Li{\v{s}}ka,~M.
  Optical Trapping of Rayleigh Particles Using a Gaussian Standing Wave.
  \emph{Opt. Commun.} \textbf{1998}, \emph{151}, 273--285\relax
\mciteBstWouldAddEndPuncttrue
\mciteSetBstMidEndSepPunct{\mcitedefaultmidpunct}
{\mcitedefaultendpunct}{\mcitedefaultseppunct}\relax
\EndOfBibitem
\bibitem[Bernatov\'{a} \latin{et~al.}(2019)Bernatov\'{a}, Donato, Je\v{z}ek,
  Pil\'{a}t, Samek, Magazz\`{u}, Marag\`{o}, Zem\'{a}nek, and
  Gucciardi]{bernatova2019wavelength}
Bernatov\'{a},~S.; Donato,~M.~G.; Je\v{z}ek,~J.; Pil\'{a}t,~Z.; Samek,~O.;
  Magazz\`{u},~A.; Marag\`{o},~O.~M.; Zem\'{a}nek,~P.; Gucciardi,~P.~G.
  Wavelength-Dependent Optical Force Aggregation of Gold Nanorods for SERS in a
  Microfluidic Chip. \emph{J. Phys. Chem. C} \textbf{2019}, \emph{123},
  5608--5615\relax
\mciteBstWouldAddEndPuncttrue
\mciteSetBstMidEndSepPunct{\mcitedefaultmidpunct}
{\mcitedefaultendpunct}{\mcitedefaultseppunct}\relax
\EndOfBibitem
\bibitem[Volpe and Volpe(2013)Volpe, and Volpe]{volpe2013simulation}
Volpe,~G.; Volpe,~G. Simulation of a Brownian Particle in an Optical Trap.
  \emph{Am. J. Phys.} \textbf{2013}, \emph{81}, 224--230\relax
\mciteBstWouldAddEndPuncttrue
\mciteSetBstMidEndSepPunct{\mcitedefaultmidpunct}
{\mcitedefaultendpunct}{\mcitedefaultseppunct}\relax
\EndOfBibitem
\bibitem[Langevin(1908)]{langevin1908theorie}
Langevin,~P. Sur la Th{\'e}orie du Mouvement Brownien. \emph{Comptes Rendus
  Acad. Sci.} \textbf{1908}, \emph{146}, 530--533\relax
\mciteBstWouldAddEndPuncttrue
\mciteSetBstMidEndSepPunct{\mcitedefaultmidpunct}
{\mcitedefaultendpunct}{\mcitedefaultseppunct}\relax
\EndOfBibitem
\bibitem[Kloeden and Platen(1999)Kloeden, and Platen]{kloeden1999numerical}
Kloeden,~P.~E.; Platen,~E. \emph{Numerical Solution of Stochastic Differential
  Equations}; Springer Verlag: Heidelberg, Germany, 1999\relax
\mciteBstWouldAddEndPuncttrue
\mciteSetBstMidEndSepPunct{\mcitedefaultmidpunct}
{\mcitedefaultendpunct}{\mcitedefaultseppunct}\relax
\EndOfBibitem
\bibitem[Callegari and Volpe(2019)Callegari, and Volpe]{callegari2019numerical}
Callegari,~A.; Volpe,~G. \emph{Flowing Matter}; Springer, 2019; pp
  211--238\relax
\mciteBstWouldAddEndPuncttrue
\mciteSetBstMidEndSepPunct{\mcitedefaultmidpunct}
{\mcitedefaultendpunct}{\mcitedefaultseppunct}\relax
\EndOfBibitem
\bibitem[Salomon \latin{et~al.}(1987)Salomon, Dalibard, Aspect, Metcalf, and
  Cohen-Tannoudji]{salomon1987channeling}
Salomon,~C.; Dalibard,~J.; Aspect,~A.; Metcalf,~H.; Cohen-Tannoudji,~C.
  Channeling Atoms in a Laser Standing Wave. \emph{Physical Review Letters}
  \textbf{1987}, \emph{59}, 1659\relax
\mciteBstWouldAddEndPuncttrue
\mciteSetBstMidEndSepPunct{\mcitedefaultmidpunct}
{\mcitedefaultendpunct}{\mcitedefaultseppunct}\relax
\EndOfBibitem
\bibitem[Meschede and Metcalf(2003)Meschede, and Metcalf]{meschede2003atomic}
Meschede,~D.; Metcalf,~H. Atomic Nanofabrication: Atomic Deposition and
  Lithography by Laser and Magnetic Forces. \emph{Journal of Physics D: Applied
  Physics} \textbf{2003}, \emph{36}, R17\relax
\mciteBstWouldAddEndPuncttrue
\mciteSetBstMidEndSepPunct{\mcitedefaultmidpunct}
{\mcitedefaultendpunct}{\mcitedefaultseppunct}\relax
\EndOfBibitem
\end{mcitethebibliography}
\begin{tocentry}
\begin{center}
\includegraphics[height=1.375in]{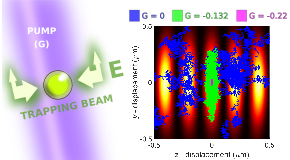}\\
\rule{0cm}{.5cm}For Table of Contents Use Only\\
\rule{0cm}{.5cm}{\bf Manuscript title:} Gain-Assisted Optomechanical Position Locking of Metal/Dielectric Nanoshells in Optical Potentials\\
\rule{0cm}{.5cm}{\bf Authors:} Paolo Polimeno, Francesco Patti, Melissa Infusino, Jonathan S\'anchez Maria A. Iat\`i, Rosalba Saija, Giovanni Volpe, Onofrio M. Marag\`o and Alessandro Veltri\\
\rule{0cm}{.5cm}{\bf Synopsis:} Gain-assisted optical trapping of dye-doped silver nanoshells is
modelled in standing wave optical potentials.
\end{center}
\end{tocentry}

\end{document}